\documentclass[12pt]{article}

\usepackage[margin=1in]{geometry}

\usepackage{setspace}

\usepackage[draft]{graphicx}

\usepackage{enumitem}

\usepackage{microtype}
\usepackage{subfigure}
\usepackage{booktabs}
\usepackage{comment}
\usepackage{wrapfig}
\usepackage{arydshln}
\usepackage{enumitem}
\usepackage{multirow}
\usepackage{url}
\usepackage{natbib}
\usepackage{authblk}

\RequirePackage[colorlinks,citecolor=blue,linkcolor=blue,urlcolor=blue,pagebackref]{hyperref}

\usepackage{amsmath}
\usepackage{amssymb}
\usepackage{mathtools}
\usepackage{amsfonts}
\usepackage{bm}
\usepackage{bbm}

\usepackage{algorithm}
\usepackage{algorithmic}

\def\eqref#1{equation~\ref{#1}}

\def\floor#1{\lfloor #1 \rfloor}
\def\1{\bm{1}}

\def\rmd{{\mathrm{d}}}

\def\rmp{{\mathrm{p}}}

\DeclareMathAlphabet{\mathsfit}{\encodingdefault}{\sfdefault}{m}{sl}
\SetMathAlphabet{\mathsfit}{bold}{\encodingdefault}{\sfdefault}{bx}{n}

\def\calB{{\mathcal{B}}}

\def\calD{{\mathcal{D}}}

\def\calI{{\mathcal{I}}}

\def\calN{{\mathcal{N}}}

\def\calT{{\mathcal{T}}}

\def\calX{{\mathcal{X}}}
\def\calY{{\mathcal{Y}}}

\def\bbE{{\mathbb{E}}}

\def\bbR{{\mathbb{R}}}

\DeclareMathOperator*{\argmin}{arg\,min}

\newcommand{\p}[1]{\left(#1\right)}
\newcommand{\sqb}[1]{\left[#1\right]}

\newcommand{\Bigp}[1]{\Big(#1\Big)}
\newcommand{\Bigsqb}[1]{\Big[#1\Big]}

\newcommand{\Biggp}[1]{\Bigg(#1\Bigg)}
\newcommand{\Biggsqb}[1]{\Bigg[#1\Bigg]}

\newcommand{\bigp}[1]{\big(#1\big)}
\newcommand{\bigsqb}[1]{\big[#1\big]}
\newcommand{\bigcb}[1]{\big\{#1\big\}}

\usepackage{amsthm}

\theoremstyle{plain}

\newtheorem{theorem}{Theorem}[section]
\newtheorem{lemma}[theorem]{Lemma}

\newtheorem{corollary}[theorem]{Corollary}

\newtheorem{definition}[theorem]{Definition}

\usepackage[textsize=tiny]{todonotes}

\renewcommand{\eqref}[1]{(\ref{#1})}

\usepackage{xcolor}
\newcount\Comments  
\newcommand{\kibitz}[2]{\ifnum\Comments=1\textcolor{#1}{#2}\fi}

\usepackage[capitalize,noabbrev]{cleveref}

\allowdisplaybreaks

\title{Debiased Nonparametric Regression for Statistical Inference and Distributionally Robustness}

\author{Masahiro Kato\thanks{Email: \texttt{mkato-csecon@g.ecc.u-tokyo.ac.jp}}$\,$}

\affil{The University of Tokyo}

\date{\today}

\begin{document}

\maketitle

\begin{abstract}
This study proposes a debiasing method for smooth nonparametric estimators. While machine learning techniques such as random forests and neural networks have demonstrated strong predictive performance, their theoretical properties remain relatively underexplored. In particular, many modern algorithms lack guarantees of pointwise and uniform risk convergence, as well as asymptotic normality. These properties are essential for statistical inference and robust estimation and have been well-established for classical methods such as Nadaraya-Watson regression. To ensure these properties for various nonparametric regression estimators, we introduce a model-free debiasing method. By incorporating a correction term that estimates the conditional expected residual of the original estimator, or equivalently, its estimation error, into the initial nonparametric regression estimator, we obtain a debiased estimator that satisfies pointwise and uniform risk convergence, along with asymptotic normality, under mild smoothness conditions. These properties facilitate statistical inference and enhance robustness to covariate shift, making the method broadly applicable to a wide range of nonparametric regression problems.
\end{abstract}

\clearpage

\section{Introduction}
This study investigates the problem of nonparametric regression. While modern machine learning methods enable the estimation of complex regression functions, they often lack the theoretical guarantees that traditional estimators possess, such as pointwise and uniform mean squared error (MSE) convergence, as well as asymptotic normality. Asymptotic normality is crucial for statistical inference, whereas uniform risk convergence plays an important role in prediction under covariate shift. In this study, we propose a debiasing method for regression estimators and show that, under mild smoothness conditions, they achieve asymptotic normality and uniform convergence.

Here, we formulate the problem. Let $X \in \calX$ be a $d$-dimensional covariate and $Y \in \calY$ be a target variable, where $\calX \subseteq \bbR^d$ and $\calY \subset \bbR$ represent the covariate and target spaces, respectively. Let $P$ denote the joint distribution of $(X, Y)$, and define the regression function under $P$ as
\[
f_P(X) = \bbE_P\bigsqb{Y \mid X},
\]
where $\bbE_P\bigsqb{\cdot}$ represents the expectation operator with respect to $P$. We assume access to observations $\bigcb{(X_i, Y_i)}_{i=1}^n$, where $(X_i, Y_i)$ are independent and identically distributed samples from a true distribution $P_0$. The true regression function is denoted by $f_0 = f_{P_0}$. Our objective is to estimate $f_0$ from the observations $\bigcb{(X_i, Y_i)}_{i=1}^n$, aiming for pointwise and uniform risk convergence, as well as asymptotic normality.

\paragraph{Notation.} For a real-valued vector $z$, let $\|z\|$ denote its Euclidean norm. For a measurable function $h\colon \calX \to \bbR$, let $\|h\|_\infty \coloneqq \sup_{x \in \calX} |h(x)|$ denote the sup-norm, and $\|h\|_2 \coloneqq \sqrt{\bbE\sqb{h(X)^2}}$ denote the $L^2$ norm.

\subsection{Content of this study}
In Section~\ref{sec:debiased_estimator}, we define our debiased estimator for nonparametric regression. Our proposed estimator consists of the following three steps: (1) estimate the regression function $f_0$ using a nonparametric regression method; (2) estimate the expected conditional residual, or equivalently, the estimation error of the first-stage estimator, using local polynomial regression; and (3) add the second-stage estimator to the first-stage estimator. 

In Section~\ref{sec:local_polynomial}, we present an example of our proposed debiased estimator, applying local polynomial regression for conditional expected residual estimation.

In Section~\ref{sec:conv_analysis}, we establish that our debiased nonparametric regression estimator achieves pointwise and uniform MSE convergence if either (a) the first-stage nonparametric regression estimator and the true regression function, or (b) their difference, belongs to the H\"older class. Using pointwise MSE convergence, we also prove asymptotic normality. Furthermore, our estimator possesses the doubly robust property: if either the first-stage nonparametric regression estimator or the second-stage conditional expected residual estimator is consistent, then the resulting debiased estimator is also consistent. 

The key takeaway is that our debiased estimator ensures desirable theoretical properties, typically associated with classical methods, for a wide range of modern first-stage estimators. As long as certain smoothness conditions are met, our method effectively corrects bias in a broad class of initial estimators, equipping the resulting estimator with strong theoretical guarantees.

\subsection{Related work}
Doubly robust and debiased estimation has attracted significant attention across various fields, including statistics \citep{tsiatis2007semiparametric}, economics \citep{ChernozhukovVictor2018Dmlf}, epidemiology \citep{BangRobins2005}, and machine learning \citep{Kallus2020double}. With the advancement of machine learning regression models, debiasing methods based on doubly robust estimators have been extensively studied \citep{ChernozhukovVictor2018Dmlf}, particularly in conjunction with sample splitting techniques \citep{klaassen1987,vanderVaart2002,ZhengWenjing2011CTME}. While much of the existing literature focuses on parametric or semiparametric models, this study extends these approaches to nonparametric regression. The application of debiased machine learning to nonparametric regression is discussed in \citet{kennedy2023semiparametricdoublyrobusttargeted}, though their focus differs from ours.

Nonparametric regression in machine learning presents several challenges:
\begin{itemize}
    \item Pointwise and uniform MSE convergence is not established for many estimators, such as neural networks and random forests.
    \item Asymptotic normality remains unproven for numerous machine learning estimators.
\end{itemize}

Pointwise and uniform MSE convergence and asymptotic normality are critical in statistical analysis, motivating significant efforts to establish these properties. For instance, \citet{Wager2018} and \citet{Mourtada2020} demonstrate asymptotic normality and optimality, respectively, for their modified random forests, but not for the original random forest of \citet{Breiman2001}.

Pointwise and uniform convergence is particularly relevant for robustness under distributional shifts, such as covariate shift \citep{Shimodaira2000}. These convergence properties are rarely established for modern nonparametric estimators, whereas classical methods like local linear and series regression possess them. This gap arises in part due to the data-adaptive nature of modern methods and the frequent use of empirical process arguments, which often focus on population MSE, where expectations are taken over the covariate distribution rather than on pointwise or sup-norm MSE. \citet{SchmidtHieber2024} addresses this concern by demonstrating restricted uniform optimality for neural network regression in the one-dimensional covariate setting, but this approach does not generalize easily.

Our method shares some motivations with debiasing approaches in high-dimensional regression. For example, the Lasso estimator, while widely used, introduces a bias that decreases with sample size at a slower rate than $\sqrt{n}$, thereby hindering asymptotic normality \citep{tibshirani96regression,Buhlmann2011}. To mitigate this, \citet{vandeGeer2014} proposes incorporating bias-correction terms, leading to improved asymptotic properties such as normality and efficiency \citep{Jankova2018}.

This study generalizes our previous work on doubly robust methods for nonparametric regression discontinuity design \citep{kato2024doublyrobustregressiondiscontinuity}. Concurrently with our work, \citet{chernozhukov2024conditionalinfluencefunctions} and \citet{vanderlaan2025automaticdebiasedmachinelearning} also explore double machine learning in the context of nonparametric regression, though their emphases differ substantially from ours. For example, \citet{chernozhukov2024conditionalinfluencefunctions} develop conditional influence functions, which are related to the framework of \citet{Ichimura2022}.

\section{The debiased estimator}
\label{sec:debiased_estimator}
In this section, we define our debiased estimator in a general way. First, we randomly split the observations $\calD$ into two datasets, $\calD^{(1)}$ and $\calD^{(2)}$, such that $\calD = \calD^{(1)} \cup \calD^{(2)}$. For simplicity, let $n$ be even and set $m = n/2$. Then, for each $\ell \in \{1, 2\}$, we define 
\[
\calD^{(\ell)} \coloneqq \bigcb{(X_i, Y_i)}_{i \in \calI^{(\ell)}},
\]
where $\calI^{(\ell)}$ is the index set of $\calD^{(\ell)}$, i.e., $\calI^{(1)} \cup \calI^{(2)} = \{1, 2, \dots, n\}$.

We consider the following three-stage estimation for each point $x_0 \in \calX$ of interest:
\begin{description}
    \item[First-stage:] Estimate $f_0$ using \emph{any} smooth model, and denote this estimator by $\widehat{f}_n$.
    \item[Second-stage:] Estimate the conditional expected residual of the first-stage estimator, $\bbE\bigsqb{Y - f(X)\mid X = x_0}$, or equivalently, the estimation error $f_0(x_0) - \widehat{f}_n(x_0)$. Denote this estimator by $\widehat{b}_n(x_0)$.
    \item[Third-stage:] Sum the first- and second-stage estimators to obtain
    \begin{align}
    \label{eq:debiased}
        &\widetilde{f}_n(x_0) \coloneqq \widehat{b}_n(x_0) + \widehat{f}_n(x_0),
    \end{align}
    which serves as a debiased estimator of $f_0$.
\end{description}

The term debiased estimator refers to our objective of correcting bias arising from the first-stage estimator. In many nonparametric and machine learning methods, bias can be problematic. By subtracting the estimated residual, $\bbE\bigsqb{Y - \widehat{f}_n(X) \mid X = x_0}$, we compensate for the difference between $f_0$ and $\widehat{f}_n$, resulting in an estimator with improved convergence and inferential properties.

Our estimator \eqref{eq:debiased} is closely related to the influence function in conditional mean estimation \citep{kennedy2023semiparametricdoublyrobusttargeted}. It is also connected to the Neyman orthogonal score \citep{ChernozhukovVictor2018Dmlf}, which is essentially equivalent to the canonical gradient in one-step bias correction \citep{Schuler2024}. Notably, in conditional mean estimation, constructing the influence function is more intricate due to the conditioning. For an influence function that involves parameters represented via the conditional expected value, see \citet{Ichimura2022}. Independently and concurrently, \citet{chernozhukov2024conditionalinfluencefunctions} investigates a related topic with different motivations and methods.

\section{The debiased nonparametric regression with local polynomial conditional expected residual estimation}
\label{sec:local_polynomial}
In this section, we present an example of our debiased estimator. We employ local polynomial regression to estimate the conditional expected residual. For simplicity, we focus on the case where the covariate is one-dimensional ($d = 1$) and $\calX = [0, 1]$.

\subsection{First-stage nonparametric regression}
For the first-stage estimator $\widehat{f}_n$, any regression method can be used, provided that smoothness conditions are met. As we show later, our key assumption is that the difference $f_0 - \widehat{f}_n$ is smooth. While various notions of smoothness exist, for simplicity, we focus on smoothness in the sense of the H\"older class (Definition~\ref{def:holder}). Notably, neither $f_0$ nor $\widehat{f}_n$ needs to be smooth individually, as long as their difference $f_0 - \widehat{f}_n$ is smooth.

We emphasize that our analysis does not impose any specific requirements on $\widehat{f}_n$ beyond smoothness. The first-stage estimator $\widehat{f}_n$ may converge to $f_0$ at an arbitrarily slow rate or even be inconsistent. Nevertheless, as long as the smoothness conditions hold, our asymptotic theoretical results remain valid. However, to achieve strong finite-sample performance, a high-quality first-stage estimator is typically preferred. In practice, methods such as random forests, neural networks, or other advanced machine learning techniques can be employed, provided they satisfy the smoothness conditions to a reasonable extent.

\subsection{Second-stage conditional expected residual estimation}
In the second-stage estimation of the conditional expected residual, we employ local polynomial regression. We define the $\ell$-th order polynomial basis as follows:
\[
\rho(u) \coloneqq \begin{pmatrix}
    1 & u & u^2/2! & \cdots & u^\ell / \ell!
\end{pmatrix}^\top.
\]
In the theoretical analysis, the value of $\ell$ is chosen based on the smoothness of $f_0 - \widehat{f}_n$.

Using this polynomial basis, we define the local polynomial regression estimator as follows:
\begin{align*}
    &\widehat{\beta}_n(x) \coloneqq \begin{pmatrix}
        \widehat{\beta}_{0, n}(x) & \widehat{\beta}_{1, n}(x) & \widehat{\beta}_{2, n}(x) & \cdots & \widehat{\beta}_{\ell, n}(x)
    \end{pmatrix}^\top\\
    &\coloneqq \argmin_{\beta \in \bbR^{\ell + 1}}\sum_{i\in\calI^{(2)}}\left(Y_i - \widehat{f}_n(X_i) - \beta^\top \rho\p{\frac{X_i - x}{h_n}}\right)^2 K\p{\frac{X_i - x}{h_n}},
\end{align*}
where $K\colon \calX \to \bbR$ is a kernel function defined by
\[
K(u) \coloneqq K_h(u) \coloneqq \mathbbm{1}\sqb{\big|u\big| \leq h}.
\]

The conditional expected residual estimator is then given by the first component of $\widehat{\beta}_n(x)$:
\[
\widehat{b}_n(x) = \widehat{\beta}_{0, n}(x).
\]

\section{Convergence analysis}
\label{sec:conv_analysis}
This section presents a convergence analysis of $\widetilde{f}_n$ defined in Section~\ref{sec:local_polynomial}. We begin by defining the H\"older class.

\begin{definition}[H\"older class]
\label{def:holder}
Given an interval $\calT \subset \bbR$, and positive constants $\beta$ and $L$, the H\"older class $\Sigma(\beta, L)$ on $\calT$ is the set of $\ell = \floor{\beta}$ times differentiable functions $f \colon \calT \to \bbR$ whose $\ell$-times derivative $f^{(\ell)}$ satisfies
\[
\Big|f^{(\ell)}(x) - f^{(\ell)}(z)\Big|\leq L \big|x - z\big|^{\beta - \ell}\quad \forall x, z \in \calT.
\]
\end{definition}

\subsection{Closed-form solution}
The debiased estimator with local polynomial regression admits a closed-form solution. We define
\begin{align*}
    \widehat{\calB}_{n}(x_0) 
    &\coloneqq \frac{1}{mh_n}\sum_{i\in\calI^{(2)}}
    \rho\p{\frac{X_i - x_0}{h_n}}
    \rho\p{\frac{X_i - x_0}{h_n}}^\top
    K\p{\frac{X_i - x_0}{h_n}},\\
    \widehat{w}_{h}(x, x_0)
    &\coloneqq \frac{1}{h_n}\rho(0)^\top
    \widehat{\calB}_{n}(x_0)^{-1}
    \rho\p{\frac{x - x_0}{h_n}}
    K\p{\frac{x - x_0}{h_n}}.
\end{align*}
Then, the second-stage estimator is given by
\[
    \widehat{b}_n(x_0) 
    = \frac{1}{m}\sum_{i\in\calI^{(2)}}
    \p{Y_i - \widehat{f}_n(X_i)}
    \widehat{w}_{h}(X_i, x_0).
\]
Thus, the debiased estimator is expressed as
\[
    \widetilde{f}_n(x_0) = \frac{1}{m}\sum_{i\in\calI^{(2)}}
    \p{Y_i - \widehat{f}_n(X_i)}
    \widehat{w}_{h}(X_i, x_0) + \widehat{f}_n(x_0).
\]

\subsection{Bias and variance decomposition}
We begin by analyzing the bias and variance of $\widetilde{f}_n$, defined as
\begin{align*}
    \mathrm{Bias}(x_0) &\coloneqq \bbE\sqb{\widetilde{f}_n(x_0)} - f_0(x_0),\\
    \mathrm{Variance}(x_0) &\coloneqq \bbE\sqb{\p{\widetilde{f}_n(x_0) - \bbE[\widetilde{f}_n(x_0)]}^2}.
\end{align*}

The bias term quantifies how closely the estimator $\widetilde{f}_n$ centers around the true regression function $f_0$ at $x_0$, while the variance term measures the fluctuation of $\widetilde{f}_n$ around its own mean. Balancing bias and variance is a fundamental concern in statistical estimation. Our approach, which employs local polynomial smoothing in the second stage, aims to systematically reduce the bias introduced by the first-stage estimator while controlling variance through appropriate bandwidth selection.

To facilitate our analysis, we introduce the population version of $\widehat{\calB}_{n}(x_0)$:
\[
    \widetilde{\calB}_{n}(x_0) 
    \coloneqq \int 
    \rho\p{\frac{x - x_0}{h_n}}
    \rho\p{\frac{x - x_0}{h_n}}^\top
    K\p{\frac{x - x_0}{h_n}}
    \rmd F(x).
\]

We then establish bounds for the bias and variance terms as follows:

\begin{theorem}[Bias and variance decomposition]
\label{thm:bias_variance}
Let $s, L, C, C_1, C_2, C_3 > 0$ be constants independent of $f_0$ and $n$. Let $h_n$ be the bandwidth of the local polynomial estimator. For every $x \in \calX$, the following hold:
    \begin{itemize}
    \item The eigenvalues of $\widetilde{\calB}_{n}(x)$ are uniformly bounded above and below away from zero.
    \item There exists a constant $C > 0$ such that
    \[
    \int \mathbbm{1}\bigsqb{| x - x_0 | \leq h_n}\rmd P(x) 
    \leq C h_n,
    \]
    where $P(x)$ denotes the distribution of $x$.
    \end{itemize}
If $f_0 - \widehat{f}_n$ belongs to the H\"older class $\Sigma(s, L)$ almost surely as $n\to \infty$, then for any $\varepsilon > 0$ and for all $x_0 \in \calX$, there exists $n_0 > 0$ such that for all $n \geq n_0$, with probability at least $1 - \varepsilon$, it holds that
\begin{align*}
    \bigl|\mathrm{Bias}(x_0)\bigr| &\leq C_1h^s_n,\\
    \mathrm{Variance}(x_0) &\leq \frac{C_2}{nh_n} + C_3h^{2s}_n.
\end{align*}
\end{theorem}

The smoothness condition on $f_0 - \widehat{f}_n$ can be satisfied in various ways; see Section~\ref{sec:allyouneed} for further discussion. The above result implies that as long as the smoothness of $f_0 - \widehat{f}_n$ is well-controlled and the bandwidth $h_n$ is chosen appropriately to shrink with $n$, both bias and variance remain bounded and ultimately vanish as $n$ grows.

\subsection{Pointwise convergence of the MSE}
From Theorem~\ref{thm:bias_variance}, the bounds on bias and variance imply the convergence of the MSE, noting that
\[
\bbE\sqb{\p{\widehat{f}_n(x_0) - f_0(x_0)}^2}
= \mathrm{Bias}(x_0)^2 + \mathrm{Variance}(x_0).
\]
To minimize the MSE, we set
\[
h_n = \alpha n^{-\frac{1}{2s + 1}},
\]
where $\alpha$ is a constant independent of $f_0$ and $n$, $s$ is the smoothness parameter, and $d$ is the dimension of $X$. We then establish the pointwise MSE convergence result.

\begin{theorem}[Pointwise MSE convergence]
\label{thm:pointwise}
Let $s, L, C, \alpha > 0$ be constants independent of $f_0$ and $n$. Let 
\[
h_n = \alpha n^{-\frac{1}{2s + 1}}
\]
be the bandwidth of the local polynomial estimator. For every $x_0 \in \calX$, the following hold:
    \begin{itemize}
    \item The eigenvalues of $\widetilde{\calB}_{n}(x)$ are uniformly bounded above and below away from zero.
    \item There exists a constant $C > 0$ such that
    \[
    \int \mathbbm{1}\bigsqb{| x - x_0 | \leq h_n}\rmd P(x) 
    \leq C h_n,
    \]
    where $P(x)$ denotes the distribution of $x$.
    \end{itemize}
If $f_0 - \widehat{f}_n$ belongs to the H\"older class $\Sigma(s, L)$ almost surely as $n\to \infty$, then for all $x_0 \in \calX$, it holds that 
\[
\bbE\sqb{\p{\widetilde{f}_n(x_0) - f_0(x_0)}^2}
= O\p{n^{-\frac{s}{2s + 1}}},
\]
as $n\to\infty$. 
\end{theorem}

Theorem~\ref{thm:pointwise} can be extended to establish the convergence of the MSE over the distribution of $X$. 

\begin{corollary}[MSE over the distribution of $X$]
Suppose the conditions in Theorem~\ref{thm:pointwise} hold. Then, we have
\[
\bbE\sqb{\p{\widetilde{f}_n(X) - f_0(X)}^2} = O\p{n^{-\frac{2s}{2s + 1}}},
\]
where the expectation is taken over both $\widetilde{f}_n$ and $X$, as $n \to \infty$. 
\end{corollary}

\subsection{Pointwise asymptotic normality}
We now establish the asymptotic normality of $\widetilde{f}(x_0)$. This result follows directly from Theorem~\ref{thm:bias_variance}.

\begin{theorem}[Asymptotic normality]
\label{thm:fastconvergencerates}
Let $s, L, C > 0$ be constants independent of $f_0$ and $n$. Let 
$h_n$ be the bandwidth of the local polynomial estimator such that
$h_n \to 0$, $nh_n \to \infty$, and $n h^{2s + 1}_n \to 0$ hold as $n\to \infty$. For every $x \in \calX$, the following hold:
    \begin{itemize}
    \item The eigenvalues of $\widetilde{\calB}_{n}(x)$ are uniformly bounded above and below away from zero.
    \item There exists a constant $C > 0$ such that
    \[
    \int \mathbbm{1}\bigsqb{| x - x_0 | \leq h}\rmd P(x) 
    \leq C h_n,
    \]
    where $P(x)$ is the distribution of $x$.
    \end{itemize}
If $f_0 - \widehat{f}_n$ belongs to the H\"older class $\Sigma(s, L)$ almost surely as $n\to \infty$, then for every $x_0 \in \calX$, it holds that
\[
\sqrt{nh_n}\p{\widetilde{f}(x_0) - f_0(x_0)}
\xrightarrow{\rmd} \calN\bigp{0, V(x_0)},
\]
as $n \to \infty$, where $V(x_0) \geq 0$ is a constant depending on $x_0$ but independent of $n$.
\end{theorem}
Here, note that the following holds from the bias and variance decomposition:
\[
V(x_0) \coloneqq nh_n\mathrm{Variance}(x_0) \leq C_2 + o(h_n).
\]

The asymptotic normality result facilitates statistical inference, such as constructing confidence intervals or hypothesis tests for $f_0(x_0)$. In practice, knowing that $\sqrt{nh_n}(\widetilde{f}(x_0) - f_0(x_0))$ converges in distribution to a normal allows for classical inferential techniques, provided that the variance $V(x_0)$ can be consistently estimated. However, constructing a confidence interval can be challenging, because $s$ (the smoothness parameter) is often unknown and affects the choice of $h_n$. In real applications, data-driven methods, including cross-validation, can be employed to select $h_n$, and bootstrap or other resampling approaches can help in variance estimation.

\subsection{Uniform convergence}
Finally, we establish the uniform (sup-norm) convergence of $\widetilde{f}_n(x_0)$. Unlike mean squared convergence, which depends on the distribution of $X$, uniform convergence bounds the supremum of the estimation error over all $x \in \calX$. This property is particularly valuable for addressing the covariate shift problem \citep{Shimodaira2000,SchmidtHieber2024}, as uniform convergence remains unaffected by changes in the distribution of $X$ between training and test datasets.

\begin{theorem}[Uniform convergence]
\label{thm:uniform}
Let $s, L_K, L, C, \alpha > 0$ be constants independent of $f_0$ and $n$. Let 
\[
h_n \coloneqq \alpha\p{\frac{\log(n)}{n}}^{\frac{1}{2s + 1}}
\]
be the bandwidth of the local polynomial estimator. For every $x_0 \in \calX$, the following hold:
    \begin{itemize}
    \item The eigenvalues of $\widetilde{\calB}_{n}(x)$ are uniformly bounded above and below away from zero.
    \item There exists a constant $C > 0$ such that
    \[
    \int \mathbbm{1}\bigsqb{| x - x_0 | \leq h_n}\rmd P(x) 
    \leq C h_n,
    \]
    where $P(x)$ denotes the distribution of $x$.
    \item The noise $\varepsilon_i$ is sub-Gaussian, satisfying
    \[
    \bbE\sqb{\exp(\lambda \varepsilon_i)}
    \leq \exp\p{K^2\lambda^2}
    \quad \forall \lambda \in \bbR,
    \]
    for some constant $K > 0$.
    \item The kernel function $K$ is Lipschitz: $K \in \Sigma(1, L_K)$ on $\calX$.
\end{itemize}
If $f_0 - \widehat{f}_n$ belongs to the H\"older class $\Sigma(s, L)$ almost surely as $n\to \infty$, then the following holds:
\[
\bbE\sqb{\left\|\widetilde{f}_n - f_0\right\|^2_\infty} = O\p{\frac{\log(n)}{n}}^{\frac{2s}{2s + 1}}.
\]
\end{theorem}

Although uniform convergence ensures robust performance under distribution shifts, it has not been established for many nonparametric regression estimators derived from modern machine learning methods. One reason for this gap is their data-adaptive nature and the reliance on empirical process techniques, which primarily focus on population risk measures whose expectation is taken over the covariate distribution rather than on sup-norm bounds. For example, \citet{SchmidtHieber2024} analyzes a neural network-based estimator in a one-dimensional covariate setting, demonstrating restricted uniform optimality. In contrast, our approach is model-free and accommodates multi-dimensional covariates, offering broader applicability and greater robustness.

\subsection{Double robustness}
Our proposed debiased estimator also possesses the property of double robustness: if either $\widehat{b}_n$ or $\widehat{f}_n$ is consistent, then $\widetilde{f}_n$ remains consistent.

Recall that
\[
    \widetilde{f}_n(x_0) 
    = \frac{1}{m}\sum_{i \in \calI^{(2)}}
    \p{Y_i - \widehat{f}_n(X_i)}\widehat{w}_h(X_i, x_0) 
    + \widehat{f}_n(x_0).
\]
Suppose $\widehat{f}_n(x_0)$ converges in probability to $f^\dagger(x_0)$ and $\widehat{b}_n(x_0)$ converges in probability to $b^\dagger(x_0)$. We analyze two cases:

\begin{itemize}
    \item If $f^\dagger(x_0) = f_0(x_0)$ but $b^\dagger(x_0) \neq f_0(x_0) - f^\dagger(x_0)$, then
    \[
    \widetilde{f}_n(x_0) 
    = \frac{1}{m}\sum_{i \in \calI^{(2)}}\Bigp{Y_i - f_0(X_i)}w^\dagger(X_i, x_0) 
    + f_0(x_0) + o_p(1),
    \]
    where $w^\dagger(X_i, x_0)$ depends only on $X_i$ and $x_0$. Since the first term converges in probability to zero, it follows that $\widetilde{f}_n(x_0) \xrightarrow{\rmp} f_0(x_0)$.

    \item If $f^\dagger(x_0) \neq f_0(x_0)$ but $b^\dagger(x_0) = f_0(x_0) - f^\dagger(x_0)$, then
    \[
    \widetilde{f}_n(x_0) 
    = \p{f_0(x_0) - f^\dagger(x_0)} + f^\dagger(x_0) + o_p(1) 
    = f_0(x_0) + o_p(1).
    \]
\end{itemize}
In both cases, $\widetilde{f}_n(x_0) \xrightarrow{\rmp} f_0(x_0)$. We summarize this result as follows.

\begin{theorem}[Double robustness]
    For any $x_0 \in \calX$, if either $\widehat{b}_n(x_0)$ or $\widehat{f}_n(x_0)$ is consistent, then
    \[
    \widetilde{f}_n(x_0) \xrightarrow{\rmp} f_0(x_0)\qquad \text{as } n\to \infty.
    \]
\end{theorem}

Double robustness thus provides an additional safeguard against model misspecification or suboptimal performance of one of the estimators. In practice, this property is particularly valuable when the first-stage model $\widehat{f}_n$ fails to accurately capture the true function, but the second-stage residual model $\widehat{b}_n$ remains reliable—or vice versa.

\section{Smoothness is all you need}
\label{sec:allyouneed}
As demonstrated above, our method relies solely on the smoothness of $\widehat{f} - f_0$. This assumption holds when both $f_0$ and $\widehat{f}$ are individually smooth, but it also remains valid in cases where neither $f_0$ nor $\widehat{f}$ is smooth, as long as their difference is smooth. Neural networks satisfy this condition under the analysis in \citet{SchmidtHieber2020}.

Notably, our results require only the smoothness of the first-stage estimator (i.e., the difference $\widehat{f}_n - f_0$). This assumption significantly broadens the applicability of our method. Even when $f_0$ and $\widehat{f}_n$ are not individually smooth, our approach can still ensure MSE convergence, asymptotic normality, and uniform convergence, provided that the difference is smooth.

From a different perspective, this flexibility allows one to use state-of-the-art or specialized methods in the first stage without compromising the theoretical guarantees of the final estimator, as long as the first-stage error exhibits a level of smoothness akin to the H\"older condition. This feature is particularly useful when the true regression function $f_0$ has a complex or partially known structure, yet the difference $f_0 - \widehat{f}_n$ remains more regular due to design choices or inductive biases in $\widehat{f}_n$.

\section{Conclusion}
This study developed a debiased estimator for nonparametric regression. For any smooth regression estimator, our debiased approach ensures pointwise and uniform MSE convergence, asymptotic normality, and double robustness. The asymptotic normality facilitates statistical inference, while uniform convergence provides distributional robustness for nonparametric regression. Additionally, if either the first-stage or the second-stage nonparametric estimator is consistent, then the resulting regression estimator remains consistent, exhibiting the doubly robust property.

Notably, our estimator requires only the smoothness of the first-stage regression estimator. Specifically, if either (a) both the first-stage estimator $\widehat{f}$ and the true regression function $f_0$ are smooth, or (b) the difference $\widehat{f} - f_0$ is smooth, then the desirable properties of the debiased estimator hold. Thus, our method imposes minimal structural constraints while guaranteeing strong theoretical results, making it an attractive and flexible approach for a wide range of nonparametric regression problems.

\bibliography{Bibtex/citation,Bibtex/causalinference,Bibtex/rdd,Bibtex/semiparametric,Bibtex/nonparametric,Bibtex/highdimensional,Bibtex/experimentaldesign,Bibtex/reinforcement,Bibtex/machinelearning,Bibtex/neuralnet,Bibtex/statistics}

\bibliographystyle{tmlr}

\clearpage

\appendix

\section*{Appendix}

\section{Proofs}
This section provides the proofs in Section~\ref{sec:conv_analysis}.

\subsection{Proof of Theorem~\ref{thm:fastconvergencerates}}
Define
\begin{align*}
    &\mathrm{Bias}\p{x_0\mid \{X_i\}_{i\in\calI^{(2)}}, \widehat{f}_n} \coloneqq \bbE\sqb{\widetilde{f}_n(x_0)\mid \{X_i\}_{i\in\calI^{(2)}}, \widehat{f}_n} - f_0(x_0),\\
    &\mathrm{Variance}\p{x_0\mid \{X_i\}_{i\in\calI^{(2)}}, \widehat{f}_n} \coloneqq \bbE\sqb{\Bigp{\widetilde{f}_n(x_0) - \bbE\sqb{\widetilde{f}_n(x_0)\mid \{X_i\}_{i\in\calI^{(2)}}, \widehat{f}_n}}^2\mid \{X_i\}_{i\in\calI^{(2)}}, \widehat{f}_n},\\
    &\mathrm{Var}\p{\bbE\sqb{\widetilde{f}_n(x_0)\mid \{X_i\}_{i\in\calI^{(2)}}, \widehat{f}_n}}\\
    &\ \ \ \ \  \ \ \ \ \ \ \ \ \ \  \ \ \ \ \ \ \ \ \ \  \ \ \ \ \ \coloneqq \bbE\Bigsqb{\Bigp{\bbE\Bigsqb{\widetilde{f}_n(x_0)\mid \{X_i\}_{i\in\calI^{(2)}}, \widehat{f}_n} - \bbE\sqb{\bbE\sqb{\widetilde{f}_n(x_0)\mid \{X_i\}_{i\in\calI^{(2)}}, \widehat{f}_n}}}^2}.
\end{align*}
Here, the followings hold:
\begin{align*}
    &\mathrm{Bias}\p{x_0} = \bbE\sqb{\mathrm{Bias}\p{x_0\mid \{X_i\}_{i\in\calI^{(2)}}, \widehat{f}_n}},\\
    &\mathrm{Variance}\p{x_0} = \bbE\sqb{\mathrm{Variance}\p{x_0\mid \{X_i\}_{i\in\calI^{(2)}}, \widehat{f}_n}} + \mathrm{Var}\p{\bbE\sqb{\widetilde{f}_n(x_0)\mid \{X_i\}_{i\in\calI^{(2)}}, \widehat{f}_n}}.
\end{align*}

We aim to show that the followings hold with probability one:
\begin{align}
\label{eq:target1}
    &\mathrm{Bias}\p{x_0\mid \{X_i\}_{i\in\calI^{(2)}}, \widehat{f}_n} \leq \frac{L}{m}\sum_{i\in\calI^{(2)}}\frac{h^s_n}{\ell !}\Big|\widehat{w}_{h}(X_i, x_0)\Big|,\\
\label{eq:target2}
    &\mathrm{Variance}\p{x_0\mid \{X_i\}_{i\in\calI^{(2)}}, \widehat{f}_n} = \frac{C_2}{nh_n}  + o(1),\\
\label{eq:target3}
&\mathrm{Var}\p{\bbE\sqb{\widetilde{f}_n(x_0)\mid \{X_i\}_{i\in\calI^{(2)}}, \widehat{f}_n}} = 2C^2_1 h^{2s}.
\end{align}

If (\ref{eq:target1}) holds, then we have
\begin{align*}
    \bbE\sqb{\frac{L}{m}\sum_{i\in\calI^{(2)}}\frac{h^s_n}{\ell !}\Big|\widehat{w}_{h}(X, x_0)\Big|} &= \int L\frac{h^s_n}{\ell !}\Big|\widehat{w}_{h}(x, x_0)\Big| \rmd P(x)\rmd x\\
    &= \int L\frac{h^s_n}{\ell !} h_n \Big|\widehat{w}_{h}(x_0 + h_nu, x_0)\Big| \rmd P(x_0 + h_nu) \rmd u,
\end{align*}
where $\frac{x - x_0}{h} = u$. We have
\begin{align*}
    &\int L\frac{h^s_n}{\ell !} h_n \Big|\widehat{w}_{h}(x_0 + h_nu, x_0)\Big| \rmd P(x_0 + h_nu) \rmd u\\
    &= L\frac{h^s_n}{\ell !} h_n \int \left|\frac{1}{h_n}\rho\p{0}^\top \widehat{\calB}_{n}(x_0)^{-1} \rho\p{u}\mathbbm{1}\sqb{\|u\| \leq 1}\right| \rmd P(x_0 + h_nu) \rmd u.
\end{align*}

\subsubsection{Preliminaries}
As preliminary, we first state the following lemma about the properties of the local polynomial estimator, as well as Proposition~1.12 in \citet{Tsybakov2008} and Appendix~B.3 in \citet{Kennedy2024}. 

\begin{lemma}[From Proposition~1.12 in \citet{Tsybakov2008}]
\label{lem:polynomial}
Let $x_0$ be a real number such that $\calB_{n}(x_0) > 0$ and let $Q$ be a polynomial whose degree is less than or equal to $\ell$. Then, the LP($\ell$) weights $\widehat{w}_n(X_i, x_0)$ satisfy
    \[\sum_{i\in\calI^{(2)}} Q(X_i)\widehat{w}_{h}(X_i, x_0) = Q(x_0).\]
for any sample $(X_1,\dots, X_n)$. 
\end{lemma}

\begin{lemma}
    Let $s, L, C > 0$ be constants independent of $f_0$ and $n$. Let $h_n$ be the bandwidth of the local polynomial estimator. For every $x_0 \in \calX$, the following hold:
    \begin{itemize}
    \item The eigenvalues of $\widetilde{\calB}_{n}(x)$ are uniformly bounded above and below away from zero.
    \item There exists a constant $C > 0$ such that
    \[
    \int \mathbbm{1}\bigsqb{| x - x_0 | \leq h_n}\rmd P(x) 
    \leq C h_n,
    \]
    where $P(x)$ is the distribution of $x$.
    \end{itemize}
    For any $\varepsilon$, there exists $n_0$ such that for all $n \geq n_0$, $h \geq \frac{1}{2n}$, and $x_0 \in \mathcal{X}$, the weights $\widehat{w}_n(x, x_0)$ of the LP($\ell$) estimator satisfy the followings:
    \begin{itemize}
        \item $\sup_{x, x_0}\Big| \widehat{w}_{h}(x, x_0) \Big| \leq \frac{C_*}{nh}$ with probability $1-\varepsilon$;
        \item $\bbE\sqb{\sum_{i\in\calI^{(2)}}\Big|\widehat{w}_n(X_i, x_0)\Big|} \leq C_* + \varepsilon$;
        \item $\widehat{w}_h(x, x_0) = 0$ if $|x - x_0| > h$,
    \end{itemize}
    where the constant $C_*$ depends only on $\lambda_0$, $a_0$, and $K_{\max}$. 
\end{lemma}

\subsubsection{Bounding the bias (proof of (\ref{eq:target1}))}

\begin{proof}
We decompose the bias term as
\begin{align*}
    &\mathrm{Bias}(x_0\mid \{X_i\}_{i\in\calI^{(2)}}, \widehat{f}_n)\\
    &= \bbE\sqb{\widetilde{f}_n(x_0)\mid \{X_i\}_{i\in\calI^{(2)}}, \widehat{f}_n} - f_0(x_0)\\
    &= \bbE\sqb{\frac{1}{m}\sum_{i\in\calI^{(2)}}\Bigp{Y_i - \widehat{f}_n(X_i)}\widehat{w}_{h}(X_i, x_0) + \widehat{f}_n(x_0)\mid \{X_i\}_{i\in\calI^{(2)}}, \widehat{f}_n} - f_0(x_0).
\end{align*}

We have
\begin{align*}
    &\bbE\sqb{\frac{1}{m}\sum_{i\in\calI^{(2)}}\Bigp{Y_i - \widehat{f}_n(X_i)}\widehat{w}_{h}(X_i, x_0) + \widehat{f}_n(x_0) - f_0(x_0)\mid \{X_i\}_{i\in\calI^{(2)}}, \widehat{f}_n}\\
    &= \bbE\sqb{\frac{1}{m}\sum_{i\in\calI^{(2)}}\Bigp{f_0(X_0) - \widehat{f}_n(X_i)}\widehat{w}_{h}(X_i, x_0) + \widehat{f}_n(x_0) - f_0(x_0)\mid \{X_i\}_{i\in\calI^{(2)}}, \widehat{f}_n}.
\end{align*}
Let $\widehat{g}_n \coloneqq \widehat{f}_n - f_0$. Since $\widehat{g}_n$ belongs to the H\"older class $\Sigma(s, L)$ almost surely as $n\to \infty$, we have
\begin{align*}
    &\bbE\sqb{\frac{1}{m}\sum_{i\in\calI^{(2)}}\Bigp{\widehat{g}_n(X_i) - \widehat{g}_n(x_0)}\widehat{w}_{h}(X_i, x_0)\mid \{X_i\}_{i\in\calI^{(2)}}, \widehat{f}_n}\\
    &= \bbE\sqb{\frac{1}{m}\sum_{i\in\calI^{(2)}}\frac{\widehat{g}^{(\ell)}(x_0 + \tau_i(X_i - x_0)) - \widehat{g}^{(\ell)}(x_0)}{\ell !}\bigp{X_i - x_0}^\ell\widehat{w}_{h}(X_i, x_0)\mid \{X_i\}_{i\in\calI^{(2)}}, \widehat{f}_n}.
\end{align*}
as $n\to\infty$, where $0\leq \tau_i \leq 1$. Here, we used the Taylor expansion of $\widehat{g}$, and Lemma~\ref{lem:polynomial}. 

Therefore, the followng holds almost surely as $n\to\infty$:
\begin{align*}
    &\bbE\sqb{\frac{1}{m}\sum_{i\in\calI^{(2)}}\frac{\widehat{g}^{(\ell)}(x_0 + \tau_i(X_i - x_0)) - \widehat{g}^{(\ell)}(x_0)}{\ell !}\bigp{X_i - x_0}^\ell\widehat{w}_{h}(X_i, x_0)\mid \{X_i\}_{i\in\calI^{(2)}}, \widehat{f}_n}\\
    &\leq \frac{1}{m}\sum_{i\in\calI^{(2)}}\frac{L\big\|X_i - x_0\big\|^s}{\ell !}\Big|\widehat{w}_{h}(X_i, x_0)\Big|\\
    &= \frac{1}{m}\sum_{i\in\calI^{(2)}}\frac{L\big\|X_i - x_0\big\|^s}{\ell !}\Big|\widehat{w}_{h}(X_i, x_0)\Big|\mathbbm{1}\bigsqb{|X_i - x_0| \leq h}\\
    &\leq \frac{L}{m}\sum_{i\in\calI^{(2)}}\frac{h^s_n}{\ell !}\Big|\widehat{w}_{h}(X_i, x_0)\Big|.
\end{align*}
\end{proof}

\subsubsection{Bounding the variance: part~I  (proof of (\ref{eq:target2}))}

\begin{proof}
As $n\to\infty$, the following holds almost surely:
\begin{align*}
    &\mathrm{Variance}(x_0\mid \{X_i\}_{i\in\calI^{(2)}}, \widehat{f}_n)\\
    &= \bbE\sqb{\Bigp{\widetilde{f}_n(x_0) - f_0(x_0)}^2\mid \{X_i\}_{i\in\calI^{(2)}}, \widehat{f}_n}\\
    &= \bbE\sqb{\Bigp{\widetilde{f}_n(x_0) - \bbE\sqb{\widetilde{f}_n(x_0)\mid \{X_i\}_{i\in\calI^{(2)}}, \widehat{f}_n}}^2\mid \{X_i\}_{i\in\calI^{(2)}}, \widehat{f}_n}\\
    &= \bbE\Biggsqb{\Biggp{\sum_{i\in\calI^{(2)}}\Bigp{Y_i - \widehat{f}_n(X_i)}\widehat{w}_{h}(X_i, x_0) + \widehat{f}_n(X_i)\\
    &\ \ \ \ \ \ \ \ \ \ \ \ - \bbE\sqb{\sum_{i\in\calI^{(2)}}\Bigp{Y_i - \widehat{f}_n(X_i)}\widehat{w}_{h}(X_i, x_0) + \widehat{f}_n(X_i)\mid \{X_i\}_{i\in\calI^{(2)}}, \widehat{f}_n}}^2\mid \{X_i\}_{i\in\calI^{(2)}}, \widehat{f}_n}\\
    &= \bbE\sqb{\p{\sum_{i\in\calI^{(2)}}\xi_i\widehat{w}_{h}(X_i, x_0)}^2 \mid \{X_i\}_{i\in\calI^{(2)}}, \widehat{f}_n}\\
    &= \sum_{i\in\calI^{(2)}}\p{\widehat{w}_{h}(X_i, x_0)}^2 \bbE\bigsqb{\xi^2_i \mid \{X_i\}_{i\in\calI^{(2)}}, \widehat{f}_n}.
\end{align*}

From the conditions we assumed, as $n\to\infty$, the following holds:
\begin{align*}
    &\bbE\sqb{\mathrm{Variance}(x_0\mid \{X_i\}_{i\in\calI^{(2)}}, \widehat{f}_n)}\\
    &\leq \frac{1}{m^2}\bbE\sqb{\sum_{i\in\calI^{(2)}}\p{\widehat{w}_{h}(X, x_0)}^2 \bbE\bigsqb{\xi^2_i \mid \{X_i\}_{i\in\calI^{(2)}}, \widehat{f}_n}}\\
    &\leq \sigma^2_{\max}\frac{1}{m}\bbE\sqb{\sup_{x,x_0}\Big|\widehat{w}_{h}(x, x_0)\Big|}\bbE\sqb{\Big|\widehat{w}_{h}(X, x_0)\Big|}\\
    &\leq \frac{\sigma^2_{\max} C^2_{*}}{mh_n}  + o(1)\\
    &= \frac{C_2}{nh_n}  + o(1).
\end{align*}
\end{proof}

\subsubsection{Bounding the variance: part~II  (proof of (\ref{eq:target3}))}
\begin{proof}
As $n\to\infty$, the following holds t
\begin{align*}
    &\mathrm{Var}\p{\bbE\sqb{\widetilde{f}_n(x_0)\mid \{X_i\}_{i\in\calI^{(2)}}, \widehat{f}_n}}\\
    &= \bbE\Bigsqb{\Bigp{\bbE\Bigsqb{\widetilde{f}_n(x_0)\mid \{X_i\}_{i\in\calI^{(2)}}, \widehat{f}_n} - f_0(x_0) + f_0(x_0) - \bbE\sqb{\bbE\sqb{\widetilde{f}_n(x_0)\mid \{X_i\}_{i\in\calI^{(2)}}, \widehat{f}_n}}}^2}\\
    &= \bbE\Bigsqb{\Bigp{\mathrm{Bias}(x_0\mid \{X_i\}_{i\in\calI^{(2)}}, \widehat{f}_n) - \mathrm{Bias}(x_0)}^2}\\
    &\leq 2C^2_1 h^{2s}_n. 
\end{align*}
\end{proof}

\subsection{Proof of Theorem~\ref{thm:uniform}}
\begin{proof}
    As $n\to\infty$, the following holds almost surely:
    \begin{align*}
        &\bbE\sqb{\Big\| \widetilde{f}_n - f_0 \Big\|^2_\infty\mid \{X_i\}_{i\in\calI^{(2)}}, \widehat{f}_n}\\
        &\leq \bbE\sqb{2\Big\| \widetilde{f}_n - \bbE\sqb{\widetilde{f}_n \mid \{X_i\}_{i\in\calI^{(2)}}, \widehat{f}_n} \Big\|^2_\infty + 2\Big\| \bbE\sqb{\widetilde{f}_n \mid \{X_i\}_{i\in\calI^{(2)}}, \widehat{f}_n} - f_0 \Big\|^2_\infty \mid \{X_i\}_{i\in\calI^{(2)}}, \widehat{f}_n}\\
        &\leq 2\bbE\sqb{\Big\| \widetilde{f}_n - \bbE\sqb{\widetilde{f}_n \mid \{X_i\}_{i\in\calI^{(2)}}, \widehat{f}_n} \Big\|^2_\infty \mid \{X_i\}_{i\in\calI^{(2)}}, \widehat{f}_n} + 2\p{\sup_{x \in \calX}\Big| \mathrm{Bias}(x_0\mid \{X_i\}_{i\in\calI^{(2)}}, \widehat{f}_n) \Big|_\infty}^2\\
        &\leq 2\bbE\sqb{\Big\| \widetilde{f}_n - \bbE\sqb{\widetilde{f}_n \mid \{X_i\}_{i\in\calI^{(2)}}, \widehat{f}_n} \Big\|_\infty \mid \{X_i\}_{i\in\calI^{(2)}}, \widehat{f}_n} + 2C^2_1h^{2s}_n.
    \end{align*}

    Here, note that 
    \begin{align*}
        &\bbE\sqb{\Big\| \widetilde{f}_n - \bbE\sqb{\widetilde{f}_n \mid \{X_i\}_{i\in\calI^{(2)}}, \widehat{f}_n} \Big\|_\infty \mid \{X_i\}_{i\in\calI^{(2)}}, \widehat{f}_n}\\
        &= \bbE\sqb{\sup_{x_0 \in \calX}\left| \sum_{i\in\calI^{(2)}} \xi_i\widehat{w}_{h}(X_i, x_0)\right|^2 \mid \{X_i\}_{i\in\calI^{(2)}}, \widehat{f}_n}.
    \end{align*}

    Hereafter, the proof step is the same as the one of Theorem~1.8 of \citet{Tsybakov2008}. We use the condition that $\varepsilon_i$ is sub-Gaussian with Corollary~1.3 and Lemma~1.6 in \citet{Tsybakov2008}. In the original Corollary~1.3 and Lemma~1.6, \citet{Tsybakov2008} assumes that $\varepsilon_i$ is Gaussian, but we can generalize the results for sub-Gaussian random variables with minor modifications.

    Finally, we have
    \begin{align*}
        \bbE\sqb{\Big\| \widetilde{f}_n - \bbE\sqb{\widetilde{f}_n \mid \{X_i\}_{i\in\calI^{(2)}}, \widehat{f}_n} \Big\|_\infty \mid \{X_i\}_{i\in\calI^{(2)}}, \widehat{f}_n} \leq C_3\frac{\log(n)}{nh_n},
    \end{align*}
    where $C_3 > 0$ is a constant independent of $f$ and $n$. 

    As $n\to\infty$, the following holds almost surely:
    \begin{align*}
        \bbE\sqb{\Big\| \widetilde{f}_n - f_0 \Big\|^2_\infty\mid \{X_i\}_{i\in\calI^{(2)}}, \widehat{f}_n} \leq C_3\frac{\log(n)}{nh} + 2C^2_1h^{2s}_n.
    \end{align*}

    By choosing the bandwidth as 
    \[h_n = \alpha\p{\frac{\log(n)}{n}}^{\frac{1}{2s + 1}},\]
    we complete the proof. 
\end{proof}

\end{document}